\documentclass[12pt]{article}
\usepackage{geometry}
\usepackage{amssymb}
\usepackage{amsmath}
\usepackage{bbm}
\usepackage{MnSymbol}
\usepackage{hyperref}
\usepackage{braket}
\usepackage{cite}
\usepackage{subcaption,tikz}
\usetikzlibrary{arrows,decorations.markings,decorations.pathreplacing}

\newcommand{\be}{\begin{equation}}
\newcommand{\ee}{\end{equation}}
\newcommand{\bea}{\begin{eqnarray}}
\newcommand{\eea}{\end{eqnarray}}

\newcommand{\mcL}{\mathcal{L}}

\newcommand{\btr}{\blacktriangleright}
\newcommand{\btl}{\blacktriangleleft}

\newcommand{\Hom}{\operatorname{Hom}}

\newcommand{\mcO}{\mathcal{O}}

\tikzset{
	partial ellipse/.style args={#1:#2:#3}{
		insert path={+ (#1:#3) arc (#1:#2:#3)}
	}
}

\numberwithin{equation}{section}

\begin{document}

\begin{center}
	
	{\large\bf Mixed Symmetries of SPT Phases}

	\vspace*{0.2in}
	
	Thomas Vandermeulen
	
	\vspace*{0.2in}
	
	{\begin{tabular}{l}
			George P.~and Cynthia W.~Mitchell Institute\\
			for Fundamental Physics and Astronomy\\
			Texas A\&M University\\
			College Station, TX 77843 \end{tabular}}
	
	\vspace*{0.2in}
	
	{\tt tvand@tamu.edu}
	
\end{center}
\pagenumbering{gobble}
\baselineskip=18pt

Symmetry Protected Topological (SPT) phases describe trivially-acting symmetries.  We argue that a symmetry-based description of SPT phases ought to include the topological twist fields associated to the symmetry.  Doing so allows us to predict the results of gauging part or all of the symmetries of these theories.

\newpage
\tableofcontents

\newpage

\section{Introduction}
\label{sec:intro}
\pagenumbering{arabic}

A Symmetry Protected Topological (SPT) phase of matter is, loosely speaking, a field theory that is trivial up to symmetry \cite{Gu_2009,Chen_2013,Kong_2020}.  The typical definition of an SPT for a symmetry $G$ is a $G$-symmetric invertible field theory.  That is, a theory with $G$ symmetry having unit partition function on any worldsheet (see e.g.~\cite[definition 2.17]{Kong_2020}, \cite[definition 4.5]{bhardwaj2023lectures}).

As emphasized in \cite{vandermeulen2023perspectives} and \cite{bhardwaj2023lectures}, the symmetry in an SPT is trivially-acting.  This does not, however, mean that its associated topological defects are trivial.  \cite{Robbins_2023} examines the topological defects associated to trivially-acting 0-form symmetries in (1+1)d, and finds that such symmetries are naturally given by a mix of line and point operators.  Said another way, a trivially-acting 0-form symmetry is a mix of 0-form and 1-form symmetries.

In this note we would like to extend this picture to higher-form symmetries of varying dimensions.  In order to motivate this pursuit, we will adopt a non-traditional definition of SPTs which more directly leverages the trivial nature of their symmetry:
\begin{center}
\fbox{\begin{minipage}{30em}
A Symmetry Protected Topological (SPT) phase associated to a $p$-form symmetry $G_{[p]}$ differs from the trivial theory by the presence of topological defects implementing a trivially-acting $G$ symmetry.
\end{minipage}}
\end{center}

The remainder of this work will endeavor to make the above definition more explicit.  Often the discussion of SPTs goes hand in hand with that of the topological twists which classify distinct SPTs exhibiting the same symmetry.  Such a treatment naturally relates to the subject of anomalies, as the admissible topological twists for an SPT classify possible 't Hooft anomalies of its boundary theories (see e.g.~\cite{Wen_2013,kapustin2014anomalies,kapustin2014symmetry,Yonekura_2019,kaidi2023symmetry}).  In this note, however, as we are concerned purely with the symmetry properties of the SPTs themselves, we will find it sufficient to disregard such twists and focus purely on the class of SPTs corresponding to non-anomalous boundary theories.  This ensures we can consistently speak of gauging bulk/boundary pairs, and does not affect the topological operators present in the theories in question (at least before gauging).  More specifically, we focus on SPTs corresponding to finite abelian bosonic symmetries.  While this may seem overly restrictive, it will yield sufficiently interesting structures when we examine the associated SPTs from the topological operator point of view.

\section{Operators in SPTs}

Let us begin with a review of this story in 2d theories.  There, 0-form symmetries are controlled by topological defect lines (TDLs).  These lines can end on local operators known as twist fields, which generically are not themselves topological.  These twist fields are bound to the endpoint of their attached TDLs -- they cannot stand freely (so are not `genuine' local operators in the sense of \cite{Kapustin_2014}).  If the symmetry in question is gaugable, in the gauged theory the twist fields become genuine, stand-alone local operators.

When will the twist fields themselves be topological?  This occurs exactly when the symmetry in question acts non-faithfully on fields on the theory.  The suggestion of \cite{Robbins_2023} is that in this case, one ought to regard the twist fields as part of the global symmetry of the theory.  This description of the topological operators in a 2d 0-form SPT matches the one given in \cite{Bhardwaj_2022}.

\begin{figure}
	\begin{subfigure}{0.5\textwidth}
		\centering
		\begin{tikzpicture}
			\draw[thick,->] (0,0) [partial ellipse=270:90:0.75cm and 0.75cm];
			\draw[thick,->] (0,0) [partial ellipse=270:-90:0.75cm and 0.75cm];
			\filldraw[black] (0,0) circle (2pt);
			\node at (0,0.3) {$\mcO$};
			\node at (-0.75,0.75) {$\mcL_g$};
		\end{tikzpicture}
		\caption{}
		\label{fig:trivaction1}
	\end{subfigure}
	\begin{subfigure}{0.5\textwidth}
		\centering
		\begin{tikzpicture}
			\draw[thick,->] (0,0) [partial ellipse=270:90:0.75cm and 0.75cm];
			\draw[thick,->] (0,0) [partial ellipse=270:-90:0.75cm and 0.75cm];
			\filldraw[black] (0,0) circle (2pt);
			\node at (0,0.3) {$\mcO$};
			\node at (-0.75,0.75) {$\mcL_g$};
			\filldraw[black] (0.75,0) circle (2pt);
			\node at (1.1,0) {$\sigma_1$};
		\end{tikzpicture}
		\caption{}
		\label{fig:trivaction2}
	\end{subfigure}
	\begin{subfigure}{0.5\textwidth}
		\centering
		\begin{tikzpicture}
			\draw[thick] (0,0) [partial ellipse=180:45:0.75cm and 0.75cm];
			\draw[thick,->] (0,0) [partial ellipse=-45:-180:0.75cm and 0.75cm];
			\draw[thick,dashed,->] (0,0) [partial ellipse=45:0:0.75cm and 0.75cm];
			\draw[thick,dashed] (0,0) [partial ellipse=0:-45:0.75cm and 0.75cm];
			\filldraw[black] (0,0) circle (2pt);
			\node at (0,0.3) {$\mcO$};
			\node at (-0.75,0.75) {$\mcL_g$};
			\filldraw[black] (0.54,0.54) circle (2pt);
			\filldraw[black] (0.54,-0.54) circle (2pt);
			\node at (0.9,0.9) {$\sigma_{g^{-1}}$};
			\node at (1.1,0) {$\mcL_1$};
			\node at (0.85,-0.9) {$\sigma_g$};
		\end{tikzpicture}
		\caption{}
		\label{fig:trivaction3}
	\end{subfigure}
	\begin{subfigure}{0.5\textwidth}
		\centering
		\begin{tikzpicture}
			\draw[thick] (0,0) [partial ellipse=180:135:0.75cm and 0.75cm];
			\draw[thick,->] (0,0) [partial ellipse=-135:-180:0.75cm and 0.75cm];
			\draw[thick,dashed,->] (0,0) [partial ellipse=135:0:0.75cm and 0.75cm];
			\draw[thick,dashed] (0,0) [partial ellipse=0:-135:0.75cm and 0.75cm];
			\filldraw[black] (0,0) circle (2pt);
			\node at (0,0.3) {$\mcO$};
			\node at (-1.1,0) {$\mcL_g$};
			\filldraw[black] (-0.54,0.54) circle (2pt);
			\filldraw[black] (-0.54,-0.54) circle (2pt);
			\node at (-0.85,0.85) {$\sigma_{g^{-1}}$};
			\node at (1.1,0) {$\mcL_1$};
			\node at (-0.85,-0.9) {$\sigma_g$};
		\end{tikzpicture}
		\caption{}
		\label{fig:trivaction4}
	\end{subfigure}
	\begin{subfigure}{\textwidth}
		\centering
		\begin{tikzpicture}
			\draw[thick,dashed,->] (0,0) [partial ellipse=270:90:0.75cm and 0.75cm];
			\draw[thick,dashed,->] (0,0) [partial ellipse=270:-90:0.75cm and 0.75cm];
			\filldraw[black] (0,0) circle (2pt);
			\node at (0,0.3) {$\mcO$};
			\filldraw[black] (-0.75,0) circle (2pt);
			\node at (-1.1,0) {$\sigma_1$};
		\end{tikzpicture}
		\caption{}
		\label{fig:trivaction5}
	\end{subfigure}
	\caption{The presence of topological twist fields allows us to `unwrap' a $G$ line.}
	\label{fig:trivaction}
\end{figure}
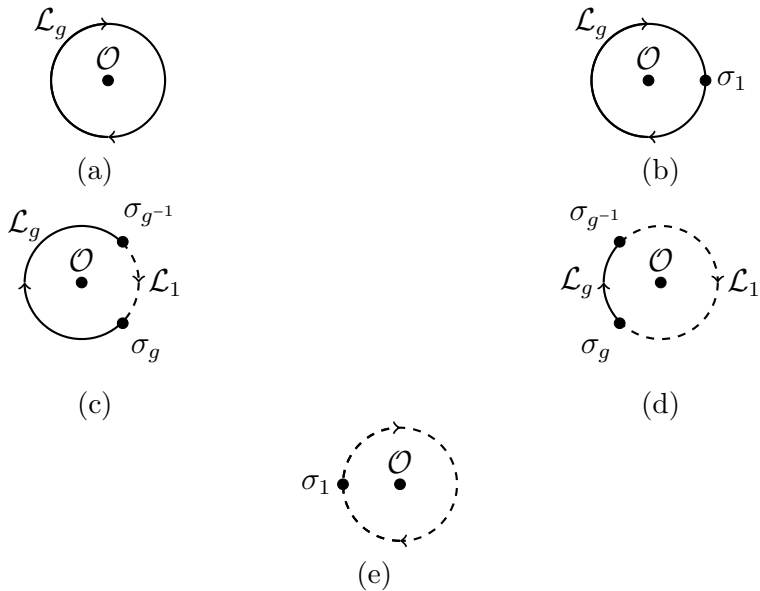

Figure~\ref{fig:trivaction} illustrates how the topological twist fields implement a trivially-acting symmetry $G$.  Given a TDL $\mcL_g$ labeled by $g\in G$, we can insert a local identity operator $\sigma_1$ and break it into an inverse pair $\sigma_g$ and $\sigma_{g^{-1}}$, such that the new adjoining line is the identity TDL.  We can then shrink $\mcL_g$ down and fuse the local operators again, resulting in the removal of all non-identity operators from $\mcL_g\cdot\mcO$.  Again, this is possible only because there exist topological junctions between $\mcL_g$ and $\mcL_1$, which would not be the case for an effectively-acting symmetry.

Let us now focus on the case of a totally trivially-acting 0-form symmetry $G$, which ought to be the symmetry of a $G$-SPT.  As argued above, one should regard $G$ as being given by a mix of topological line and point operators, each labeled by $G$.  We would like to regard this total symmetry $\Gamma_G$ of the system as being given by the short exact sequence
\be
\label{01ext}
1\to G_{[0]}\to \Gamma_G\to G_{[1]}\to 1.
\ee
That is, we will regard our total symmetry as an extension of a 1-form $G$ symmetry by a 0-form $G$ symmetry.

Let us assign some notation.  We will let
\be
\label{attach}
G_{[p]}\btr G_{[p+1]}
\ee
indicate that the operators for $G_{[p+1]}$ are `attached' to those of $G_{[p]}$, meaning that the $G_{[p+1]}$ operators are elements of the junction Hilbert space at the codimension 1 intersection of two $G_{[p]}$ operators.  Our definition of an SPT can then be more precisely stated as

\begin{center}
	\fbox{\begin{minipage}{30em}
			A Symmetry Protected Topological (SPT) phase associated to a $p$-form symmetry $G_{[p]}$ in $d\ge 2$ spacetime dimensions differs from the trivial theory by the presence of topological defects corresponding to 
			\[
			G_{[p]}\btr G_{[p+1]},
			\]
			with $0\le p\le d-2$.\\
			
			That is, there exist in the theory dimension $d-(p+2)$ topological defects attached to dimension $d-(p+1)$ topological defects, both labeled by $G$.
	\end{minipage}}
\end{center}

\section{Bulk/Boundary and Gauging}

A $d$-dimensional SPT can naturally serve as a boundary theory for a $(d+1)$-dimensional SPT, given suitable boundary conditions.  Quite simply, fixing the operators in the $(d+1)$-dimensional bulk SPT to intersect the boundary in codimension 1 leads to operators one dimension lower on the boundary, exactly as we would expect from a $d$-SPT.  As a concrete example, a 0-form SPT in 3 dimensions contains topological surfaces with attached TDLs.  When these operators intersect a two-dimensional boundary in codimension 1, the boundary theory will have TDLs ending on attached topological point operators.  This is precisely the operator content of a 2d 0-form SPT, as claimed.

Now we can ask what happens upon gauging.  We know to expect gauging a $p$-form symmetry to produce a dual $d-(p+2)$-form symmetry.  In the setup above, both the bulk and boundary have the symmetry $G_{[p]}\btr G_{[p+1]}$.  Let us gauge the $G_{[p]}$ symmetry.  Gauging will convert the non-trivial extension class\footnote{Following \cite[A.3]{Tachikawa_2020}, extensions of the form (\ref{attach}) ought to be classified by $H^{p+2}(K(G,p+2),G) = \Hom(G,G)$ (where $K(G,m)$ is the $m$th Eilenberg-MacLane space associated to $G$).  Recall that we are restricting here to abelian $G$.} to a mixed anomaly \cite{Tachikawa_2020}, with the resulting symmetry of the gauged theory given by the direct product
\be
\hat{G}_{[d-(p+2)]} \times G_{[p+1]}.
\ee

While the bulk and boundary once again appear to have the same dual symmetry -- $\hat{G}_{[d-(p+2)]}$ -- note that the factor of $d$ implies that our boundary conditions must have changed.  This symmetry is implemented by $(p+1)$-dimensional defects, with the important part being that this expression is independent of $d$.  This means that, in order to match up, the bulk $(p+1)$-dimensional defects must be fully localizable to the boundary.  This is a general behavior; the dual of a symmetry whose defects intersect the boundary in codimension 1 will have localizable defects, and vice-versa.

\begin{figure}[h]
\centering
\begin{tabular}{c c | c c | c}
Boundary & Op.~Dim. & Bulk & Op.~Dim. & Boundary Conditions\\
\hline
$G_{[0]}\btr G_{[1]}$ & $1\btr 0$ & $G_{[0]}\btr G_{[1]}$ & $2\btr 1$ & $\perp \btr \perp$\\
$\hat{G}_{[0]}\times G_{[1]}$ & $1\times 0$ & $\hat{G}_{[1]}\times G_{[1]}$ & $1\times 1$ & $\parallel\times\perp$\\
$\hat{G}_{[0]}\btl\hat{G}_{[-1]}$ & $1\btl 2$ & $\hat{G}_{[1]}\btl\hat{G}_{[0]}$ & $1\btl 2$ & $\parallel\btl\parallel$\\
\end{tabular}
\caption{Successive gaugings of a 3d 0-form SPT with 2d boundary.  At each step of the gauging we list the symmetry and dimensions of operators that implement those symmetries on both bulk and boundary, along with the boundary conditions.}
\label{fig:3d2d}
\end{figure}

Again, a 3d bulk with a 2d boundary provides an easily visualizable example, the results of which are tabulated in Figure~\ref{fig:3d2d}.  A 0-form SPT in three spacetime dimensions has symmetry $G_{[0]}\btr G_{[1]}$, implemented by surface operators with attached lines.  These intersect the boundary in codimension 1, resulting in lines with attached points.  Gauging the $G_{[0]}$ symmetry produces line operators in both 3d and 2d, meaning that the 3d lines must be localizble to the boundary.  The resulting 3d theory has symmetry $\hat{G}_{[1]}\times G_{[1]}$ -- this theory of a gauged trivially-acting $G$ symmetry often goes by the name `$G$ gauge theory', and the line operators are the Wilson and 't Hooft lines for the electric and magnetic symmetries.  The gauged boundary theory has a 1-form symmetry, which is a $(d-1)$-form symmetry, indicating a ground-state degeneracy.  This operator content is consistent with the expectation that gauging the 0-form symmetry of a 0-form SPT produces a TQFT of Dijkgraaf-Witten type -- in 2d such theories are known to decompose, while in 3d their operator content consists of anyons (corresponding to the 1-form symmetries).

As stated earlier, there is a mixed anomaly between $\hat{G}_{[1]}$ and $G_{[1]}$, but we are no longer prevented from gauging $G_{[1]}$ (as the associated lines have become genuine operators in the gauged theory).  Note that, after this first gauging, the boundary theory is a 2d theory with 1-form symmetry, hence it contains multiple local topological operators and thus exhibits ground state degeneracy, i.e.~it decomposes into a direct sum.  Let us gauge $G_{[1]}$ in both the bulk and boundary.  In the bulk we have already seen that gauging exchanges 0-form and 1-form symmetries, so we return to our original situation of surface operators with lines attached: the bulk theory has gone from a $G_{[0]}$ SPT to a $\hat{G}_{[0]}$ SPT.  On the boundary, however, we are now gauging a 1-form symmetry in 2d.  The expected dual to this is a $-1$-form symmetry controlled by 2-dimensional topological operators.  This is consistent with the bulk theory being an `ordinary' 0-form SPT because of the change in boundary conditions -- the bulk $\hat{G}$ surface operators are now localizable to the boundary, where they serve as spacetime-filling operators implementing the $-1$-form symmetry.

Through two successive gaugings we have thus gauged the entire initial $G_{[0]}\btr G_{[1]}$ symmetry on both the bulk and boundary, which would be a `full gauging' of the mixed symmetry of the 0-form SPT.  In 3d this simply produced a new 0-form SPT -- these theories map to themselves under such a procedure.  In 2d we produced something new, which one could reasonably call an SPT for a $-1$-form symmetry.  In $d$ dimensions, this would be a theory that differs from the trivial one by the inclusion of $d$-dimensional spacetime-filling defects with attached $(d-1)$-dimensional twist operators.  This allows us to extend the range of validity of the previous section's definition of SPT to include $p=-1$.

\section{SPTs in Quantum Mechanics}

So far we have discussed SPTs for $p$-form symmetries with $-1\le p\le d-2$.  Conspicuously absent is $p=d-1$, which should be its maximum attainable value -- a $(d-1)$-form symmetry is the highest-form symmetry we can have in any given dimension, as its associated topological operators are 0-dimensional (local).  Because of this, our previous characterization of trivially-acting symmetries which highlighted the operators living at codimension 1 junctions no longer applies: there are no codimension 1 junctions for local operators.

The most obvious test case here is that of quantum mechanics (QM), i.e.~$d=1$.  In QM, 0-form symmetries correspond to local operators.  What topological defects should we assign to a trivially-acting symmetry in QM?  To answer this we note that a 0-form SPT in QM should be realizable on the boundary of a 0-form SPT in 2d, with the same general setup as presented in the previous section.  That is, we have a 2d $G$-SPT with a 1d boundary.  We choose boundary conditions such that the $G$ TDLs are perpendicular to the boundary, allowing them to match up with the local $G$ operators that ought to exist in the QM SPT.

\begin{figure}[h]
\begin{subfigure}{0.5\textwidth}
	\centering
	\begin{tikzpicture}
		\draw[thin] (-1,0) -- (1,0);
		\filldraw[black] (0,0) circle (2pt);
		\node at (0,-0.5) {$\tau_g$};
		\draw[thick,->] (0,1) -- (0,0.5);
		\draw[thick] (0,0.5) -- (0,0);
		\node at (0.5,0.5) {$\mcL_g$};
		\filldraw[black] (0,1) circle (2pt);
		\node at (-0.7,1) {$\sigma_{gh^{-1}}$};
		\draw[thick,->] (0,2) -- (0,1.5);
		\draw[thick] (0,1.5) -- (0,1);
		\node at (0.5,1.5) {$\mcL_h$};
	\end{tikzpicture}
\caption{}
\label{1d2da}
\end{subfigure}
\begin{subfigure}{0.5\textwidth}
	\centering
	\begin{tikzpicture}
		\draw[thin] (-1,0) -- (1,0);
		\filldraw[black] (0,0) circle (2pt);
		\node at (0,-0.5) {$\sigma_{gh^{-1}}\cdot\tau_g=\tau_h$};
		\draw[thick,->] (0,2) -- (0,1);
		\draw[thick] (0,1) -- (0,0);
		\node at (0.5,1) {$\mcL_h$};
	\end{tikzpicture}
\caption{}
\label{1d2db}
\end{subfigure}
\caption{An operator $\sigma_{gh^{-1}}$ coming in from the bulk changes the boundary operator $\tau_g$ to $\tau_h$.}
\label{fig:1d2d}
\end{figure}
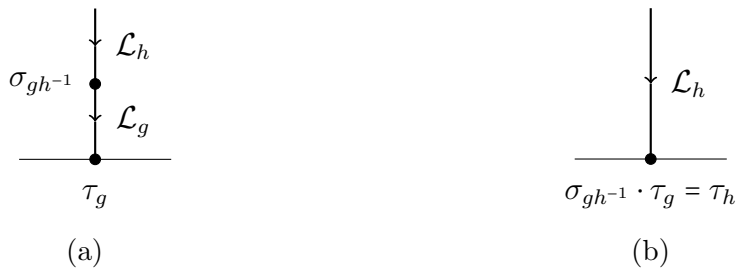

The 2d twist fields, as we well know, are local operators that live at TDL junctions.  As they have no spatial extent, there is no choice of boundary conditions for them.  We can, however, still move them to the boundary along their attached TDLs, as shown in Figure~\ref{fig:1d2d}.  There the TDLs implementing the 2d $G_{[0]}$ symmetry are written as $\mcL_g$ and the local operators implementing $G_{[1]}$ are $\sigma_g$.  The local operators implementing $G_{[0]}$ on the 1d boundary are $\tau_g$.  Going from Figure~\ref{1d2da} to Figure~\ref{1d2db}, we shrink the $\mcL_g$ line, moving $\sigma_{gh^{-1}}$ to the boundary.  The resulting boundary operator, as it is attached to the line $\mcL_h$, must be $\tau_h$.  Thus the operators $\sigma$ must also exist on the boundary and act on the $\tau$ as
\be
\sigma_h\cdot\tau_g = \tau_{h^{-1}g}.
\ee
From the point of view of the boundary theory, then, there are two sets of local topological operators: the $\tau_g$ and the $\sigma_g$.  The $\tau$ are genuine local operators, while the $\sigma$ can exist only on the $\tau$.  These clearly constitute a trivially-acting symmetry since, for any field $\varphi$ acted on by $\tau$, we have
\be
\tau_g\cdot\varphi = (\sigma_g\cdot\tau_g)\cdot\varphi = \tau_{1}\cdot\varphi=\varphi,
\ee
meaning that $\tau_g$ acts the same as $\tau_1$ (which must be the identity) on all fields in the theory.  In order for this process to make sense on the boundary alone, the $\sigma$ must constitute a set of boundary local operators, distinct from the $\tau$.  Their presence will additionally ensure that bulk and boundary symmetries match upon gauging, and that the gauged SPT is $G$-symmetric (note that we expect the resulting spontaneous symmetry breaking phase to have degenerate ground states; the $\sigma$ take care of this).

So we have another case of `attachment' of operators, this time of the same dimension.  We can thus write the symmetry of this 1d SPT as
\be
G_{[0]}\btr G_{[0]}.
\ee
We can then play the same game as in the previous section, successively gauging this bulk/boundary combination of SPTs.  The results of this exercise are summarized in Figure~\ref{fig:1d2d_table}.

\begin{figure}[h]
	\centering
	\begin{tabular}{c c | c c | c}
		Boundary & Op.~Dim. & Bulk & Op.~Dim. & Boundary Conditions\\
		\hline
		$G_{[0]}\btr G_{[0]}$ & $0\btr 0$ & $G_{[0]}\btr G_{[1]}$ & $1\btr 0$ & $\perp \btr \emptyset$\\
		$\hat{G}_{[-1]}\times G_{[0]}$ & $1\times 0$ & $\hat{G}_{[0]}\times G_{[1]}$ & $1\times 0$ & $\parallel\times\emptyset$\\
		$\hat{G}_{[-1]}\btl\hat{G}_{[-1]}$ & $1\btl 1$ & $\hat{G}_{[0]}\btl\hat{G}_{[-1]}$ & $1\btl 2$ & $\parallel\btl\emptyset$\\
	\end{tabular}
	\caption{Successive gaugings of a 2d 0-form SPT with a 1d 0-form SPT on its boundary.}
	\label{fig:1d2d_table}
\end{figure}

If we instead started out with the 2d 0-form TDLs having parallel intersection with the boundary, the 1d theory ought to have symmetry $G_{[-1]}\btr G_{[0]}$, i.e.~be a $-1$-form SPT.  Gaugings of this combination are presented in Figure~\ref{fig:1d2d_table2}.  We see that the 1d $-1$-form SPTs are self-dual under gauging of their full symmetry, as were 0-form SPTs in 3d.

\begin{figure}[h]
	\centering
	\begin{tabular}{c c | c c | c}
		Boundary & Op.~Dim. & Bulk & Op.~Dim. & Boundary Conditions\\
		\hline
		$G_{[-1]}\btr G_{[0]}$ & $1\btr 0$ & $G_{[0]}\btr G_{[1]}$ & $1\btr 0$ & $\parallel \btr \emptyset$\\
		$\hat{G}_{[0]}\times G_{[0]}$ & $0\times 0$ & $\hat{G}_{[0]}\times G_{[1]}$ & $1\times 0$ & $\perp\times\emptyset$\\
		$\hat{G}_{[0]}\btl\hat{G}_{[-1]}$ & $0\btl 1$ & $\hat{G}_{[0]}\btl\hat{G}_{[-1]}$ & $1\btl 2$ & $\perp\btl\emptyset$\\
	\end{tabular}
	\caption{Successive gaugings of a 2d 0-form SPT with a 1d $-1$-form SPT on its boundary.}
	\label{fig:1d2d_table2}
\end{figure}

The above analysis should extend to $(d-1)$-form symmetries (implemented by topological local operators) in $d$ dimensions, allowing us to once again extend our SPT definition to include the edge case $p=d-1$:

\begin{center}
\fbox{\begin{minipage}{30em}
An SPT corresponding to a $(d-1)$-form symmetry $G$ in $d$ spacetime dimensions differs from the trivial theory by the inclusion of local topological operators corresponding to
\[
G_{[d-1]} \btr G_{[d-1]}.
\]
\end{minipage}}
\end{center}

\section{Conclusion}

Many aspects of SPTs can be understood purely through analysis of their topological operators, which possess a richer structure than the usual codimension-$(p+1)$ defects which implement a $p$-form symmetry.

Here we focused on SPTs for pure (i.e.~not mixes of differing form) symmetries.  We could, for instance, ask about SPTs for 2-group symmetries (that is, the extension of a 0-form symmetry by 1-form symmetry).  In order to treat such an SPT in the language used here, one would need a topological operator description of trivially-acting 2-groups.\footnote{3d orbifolds by trivially-acting 2-groups were studied in \cite{Pantev_2022}.}  A similar sentiment holds for non-invertible symmetries.

\section{Acknowledgments}

I would like to thank Daniel Robbins and Eric Sharpe for helpful comments on a draft of this note.

\addcontentsline{toc}{section}{References}

\bibliographystyle{utphys}
\bibliography{SPTNotes}

\end{document}